# Dengue Fever in Perspective of Clustering Algorithms

Kamran Shaukat[1*], Nayyer Masood[2], Ahmed Bin Shafaat[1], Kamran Jabbar[1], Hassan Shabbir[1] and Shakir Shabbir[1]
[1]University of the Punjab, Jhelum Campus, Pakistan
[2]Mohammad Ali Jinnah University, Islamabad Campus. Pakistan

**Abstract**

Dengue fever is a disease which is transmitted and caused by Aedes Aegypti mosquitos. Dengue has become a serious health issue in all over the world especially in those countries who are situated in tropical or subtropical regions because rain is an important factor for growth and increase in the population of dengue transmitting mosquitos. For a long time, data mining algorithms have been used by the scientists for the diagnosis and prognosis of different diseases which includes dengue as well. This was a study to analyses the attack of dengue fever in different areas of district Jhelum, Pakistan in 2011. As per our knowledge, we are unaware of any kind of research study in the area of district Jhelum for diagnosis or analysis of dengue fever. According to our information, we are the first one researching and analyzing dengue fever in this specific area. Dataset was obtained from the office of Executive District Officer EDO (health) District Jhelum. We applied DBSCAN algorithm for the clustering of dengue fever. First we showed overall behavior of dengue in the district Jhelum. Then we explained dengue fever at tehsil level with the help of geographical pictures. After that we have elaborated comparison of different clustering algorithms with the help of graphs based on our dataset. Those algorithms include k-means, K-mediods, DBSCAN and OPTICS.

## Introduction

Dengue fever is an infectious disease transmitted by Aedes Aegypti mosquitos. Dengue has infected 1.2 billion people in 56 countries till 1998 and now it is present in all regions of World Health Organization WHO [1]. Another study shows that 2500 million lives of total worldwide population are at risk due to dengue fever. Dengue hemorrhagic fever and dengue fever are causing annually 500,000 and 50-100 million cases respectively. 24000 cases out of these calculations are leading to death. This study shows that dengue has increased 30 times from 1960 to 2015 [2]. Dengue fever is a serious public health issue in Pakistan due to number of increased deaths in city of Lahore and Karachi in last few years [3,4]. The first confirmed dengue attack in Pakistan was noticed in 1994. Another attack was noticed in Baluchistan in October 1995 [5]. In Pakistan, dengue has become a serious threat for public health due to unavailability of vaccine, unclean water, rapidly increasing number of dengue fever cases, highly populated areas and low quality system of sanitation and sewerage. There has been a lot of research done on diagnosis of dengue fever and a number of methods have been proposed for the classification and clustering of dengue fever in the world. G.P.Silveira et al proposed a kind of Takagi-Sugeno model for the analysis of dengue fever in Brazil. Silveira used fuzzy rule and partial differential equation for the analysis of data. The uncertainty factors discussed in their study directly relate with the number of humans which provide blood for the maturation of mosquito's eggs. Silveira considered rain factor for the increase or decrease in mosquito's population [6]. Another aspect of analysis of dengue virus (DENV) is to determine the effect of DENV on people who are living in the neighbourhood of cases infected by DENV. There has been a study on the transmission of dengue fever in highly urban Vietnam. Anders KL proved that in highly urban areas people who were living close to the dengue infected cases were in fact not infected over a period of two weeks [7]. N. Subitha et al have proposed spatial data mining techniques and algorithms for the diagnosis of dengue fever. They proposed k-means algorithm for the extraction of patterns from spatial databases. They used microscopic blood image report for the implementation of an automated system which reorganizes the dengue fever. They used microscope image report as input and then they filtered the signals and extracted feature characteristics. These features were then fed to neural networks and then classification was done using Back Propagation Network (BPN) [8]. There is another disease which is very similar to dengue fever DENV named chikungunya (CHIK) virus caused an alarming situation in tropical regions. Chikungunya disease is caused due to the mutual sharing of same vector with arbovirus and dengue virus. Fathima A. Shameem et al observed physical and clinical diagnosis of patients who were infected from chikungunya viral fever and then compared it with dengue fever. They proposed data mining process and knowledge discovery process (KDD) for the prognosis and diagnosis of arbovirus-dengue [9]. Daranee Thitiprayoonwongse et al proposed a hybrid method which combines a decision tree and fuzzy logic approach for building dengue infection disease model. They explored two classification problems which are dengue classification problem and the day of defervescence of fever (day0) detection problem. They proposed to use the knowledge obtained from the decision tree with the fuzzy logic approach. Their experiments showed that using fuzzy logic approach for dengue classification is a suitable method [10]. There has been another aspect of the analysis of dengue virus which is the adaptation of p-means algorithm and classification. María Beatríz Bernábe-Loranca et al used clustering with predetermined centroids as a registered case of dengue fever. Based on p-means algorithm a clustering algorithm has been associated and adopted by Maria et al which constructs groups where center of each group is a dengue infected case [11]. Shafique Ahmed et al showed a statistical study of those countries in which dengue fever has effected certainly with the help of annotated maps. And in Pakistan they showed a specific group among society which was least resistant against dengue fever [1]. Sadia Nasreen et al measured the frequency and distribution of dengue fever in district Faisalabad, Pakistan. They have analysed the epidemiology of disease in Faisalabad [2]. Bilal Tariq and Arjumand Z. Zaidi discovered relationship between dengue fever outbreaks and land cover/land use in Lahore. They used remote sensing data for the identification of those environmental factors who contribute in dengue fever outbreaks. Then they correlated the dengue outbreak distribution pattern with those identified environmental factors. To identify risk prone areas they have developed a geo-statistical dengue risk model by integrating demographic, land cover/use and environmental parameters with dengue cases [12].

## Critics

K-means and p-means algorithms are mostly used clustering algorithms in the domain of data mining. Both of these algorithms have a variety of applications for clustering. But they also have some disadvantages. When k-means algorithm encounters with a sufficiently large or very low value called outlier, that outlier disturbs the whole result. K-means and p-means both algorithms are sensitive to outliers. K-means algorithm does not show better results when it works with clusters of different size and density [13]. They do not cater distance. K-means algorithm is unable to overcome this problem. This problem is due to the fact that k-means and p-means algorithms use mean values as centriods.To solve this issue, we propose dbscan algorithm for the clustering of dengue fever infected cases in Jhelum district.

Dbscan was first density-based algorithm. It discovers arbitrarily shaped clusters in spatial databases with noise. Dbscan algorithm performs more efficiently than k-means while forming clusters due to the fact that dbscan algorithms uses modes as centroids. That is why dbscan shows better and more correct results than k-means. Another reason for using dbscan is that it discovers clusters of arbitrary shapes and regions. By using dbscan, clusters having high density can be separated from regions which have low density so a better visualization of dataset can be achieved and can be effectively used in decision making.

## Objective

According to any dataset achieved, one can only tell the number of victims in any specific area. Like in this case, one can say that there were 200 cases of dengue in Jhelum in 2011. But he cannot analyze that in which region dengue attacked more people? Or what is the reason of their being infected by dengue? Some of infected cases were basically from Jhelum but they worked in other cities which were main victims of dengue fever. They were being infected in other cities and then they came in Jhelum. So they behaved as 'carrier'. By using our technique one can truly analyze the attack of dengue and can make precaution in future.

The research objective is to use one of the density-based algorithms to determine the population of dengue fever infected cases in Jhelum district and in surrounding areas geographically. So that a better visualization of dengue infected cases can be achieved and be used in good decision making. Clustering techniques are used for analyzing the attack of dengue so that precautions can be made. By using this technique one can better analyze the attack of dengue fever in any area. This research includes the comparison of different clustering algorithms with the help of graphs, based on chunk of dataset.

## Methodology

District Jhelum is situated in province of Punjab, Pakistan. The location of Jhelum city is 32.9286°N longitude and 73.7314°E latitude. Dengue fever has become a serious issue in Jhelum since 2011 when a large number of dengue cases were registered in EDO's office (health). Dataset for this research paper is also taken from the office of EDO (health) Jhelum.

There are many algorithms in data mining that can be used for clustering. We are using one of the density-based algorithms i.e. dbscan algorithm. But dbscan also has some drawbacks. Like many other clustering algorithms dbscan is also sensitive to parameter values. A small difference in values of parameter may cause a very different clustering of data. To handle such kind of problems, Optics and DenClue algorithms were proposed.In our work, first we presented dengue infected cases in general and then in detailed form. We clustered the dengue infected cases on geographical regions. Firstly while representing the whole district, we clustered the dataset on the basis of tehsils (a geographical sub-unit used in Pakistan). District Jhelum is consists of four tehsils. After that, when we represented each tehsil, we clustered on the basis of villages. That is why a complete and brief clustering model of dengue fever has been achieved in Jhelum district. There were some dengue cases who did not belong to Jhelum district. They were from areas like Sara-e-Alamgir district Gujrat so we treated them as outliers. The values of parameters taken in this paper are

o   $\varepsilon = 1$ cm

o   MinPts=3

In the Figure 1 shown below, we have represented overall dengue cases in whole district Jhelum. From this fig. we can observe that tehsil of Jhelum was highly infected by dengue fever. 46 dengue infected cases out of 95 were registered from tehsil of Jhelum which shows that tehsil Jhelum was 48% infected by dengue. Tehsil of Dina had comparatively smaller but approximately same attack of dengue as in Jhelum tehsil because of its neighborhood with Jhelum tehsil. Round about 22 i.e. 23% of all cases of dengue fever were registered from tehsil Dina. We can observe from the fig. that a big cluster is representing both tehsils of Jhelum and Dina due to very small distance between them. But we will discuss each tehsil separately later. Tehsil Sohawa and tehsil Pind Dadan Khan had comparatively smaller attack of dengue fever as compared to tehsils of Jhelum or Dina and it can be observed from the (Figure 1).

Some cases were registered from surrounding areas of tehsil Sohawa but overall Sohawa city was not too much infected by dengue fever. Similarly tehsil Pind Dadan Khan was not highly infected by dengue. The reason behind it can be its larger distance from Jhelum and Dina. It is located far from Dina and Jhelum. There were some cases registered from Sara-e-Alamgir which is also joined with Jhelum but since its region is Gujrat so we considered it as an outlier. Now we will elaborate dengue behavior in each tehsil of district Jhelum.

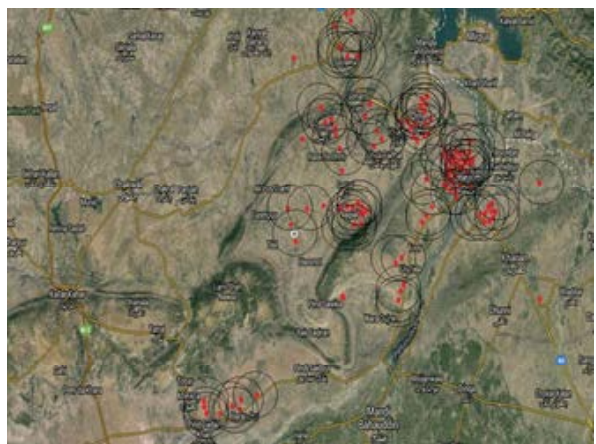

**Figure 1**: Dengue fever infected cases in district Jhelum during 2011.

### Tehsil Dina

Dina is located at 33°01′42″ latitude and 73°36′04″ longitude. From the figure 2 we can observe that tehsil Dina was infected mostly in rural areas. But there were some cases registered from Dina city also. There is a growing cluster in all over tehsil Dina which is consists of area from Chak Abdul Khaliq to the village of Mota Gharbi. A comparatively smaller cluster can be observed in the region of Rohtas Fort and Bhagaan Village. Village Dhok Bhadaal has a very small attack of dengue fever since a little cluster is made in the region of Dhok Bhadaal shown in (Figure 2).

### Tehsil Sohawa

Sohawa is situated at 32° 49' 30" latitude and 73° 45' 55" longitude. Tehsil Sohawa is consists of almost rural areas. Mostly cases of dengue fever were recorded from rural areas. But a few cases were from city too. Three clusters can be observed in the tehsil Sohawa. The larger one consists of area from Noor Poor Syedaan to the village of Kandyari and the cluster next to it is situated in the area of Diyaali and Rakha villages. The first two clusters are indirectly approachable. The third cluster is in the area of Domeli and Jodha which is at a little distance from other two clusters as shown in the (Figure 3).

### Tehsil Pind Dadan Khan

Pind Dadan khan is located at 32°34'60 latitude and 73°2'60 longitude and it contain the world's largest salt mine which is Khewra salt mine. Tehsil Pind Dadan Khan is showing four clusters in its region. First one is in the area of Khewra Salt Mine and Pind Dadan Khan City which is a dense cluster than others.

The next cluster is in Haran Pur. Next two clusters are indirectly reachable and they are in a comparatively large area consisting Dharyala Jalab and Pananwal shown in (Figure 4).

### Tehsil Jhelum

In tehsil Jhelum we can observe that mostly urban area i.e. Jhelum city was under dengue attack due to a very dense cluster in the Jhelum city. And this growing cluster is consists of area of Jhelum Cantonment to Kala Gujran.

The other clusters of tehsil Jhelum are representing rural areas like Sanghoi. Nurpur Baghan, Hunh Humwala and Chotala shown in (Figure 5).

### Graph Comparison of Algorithms

Many clustering algorithms have been proposed in the field of data mining. Every algorithm have some advantages and disadvantages as well. Now we will discuss four clustering algorithms which include k-means, k-mediods, dbscan and optics. We will elaborate a graph comparison of these algorithms based on our chunk of dataset.

### K-MEANS

K-means is one of the simplest algorithm of clustering (unsupervised learning). K-means has the advantage of its simplicity and robustness [14]. Although its worst case behavior is slow but if we keep k small and parameter values are comparatively bigger, then k-means performs faster computations than hierarchical clustering [13]. K denotes the number of clusters we want to form. In this case we want to make three clusters so we keep k=3. But k-means also has some disadvantages. When clusters are global then k-means does not work well with them. K-means cannot handle outlier and noisy data because it do not cater distance. When k-means have to work with non-linear dataset then it fails [15]. The prediction of k-value becomes very difficult when numbers of clusters are fixed shown in (Figure 6) [16].

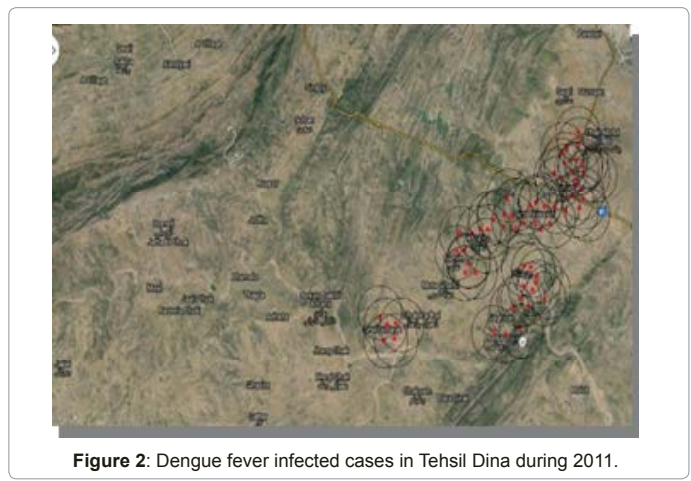

**Figure 2**: Dengue fever infected cases in Tehsil Dina during 2011.

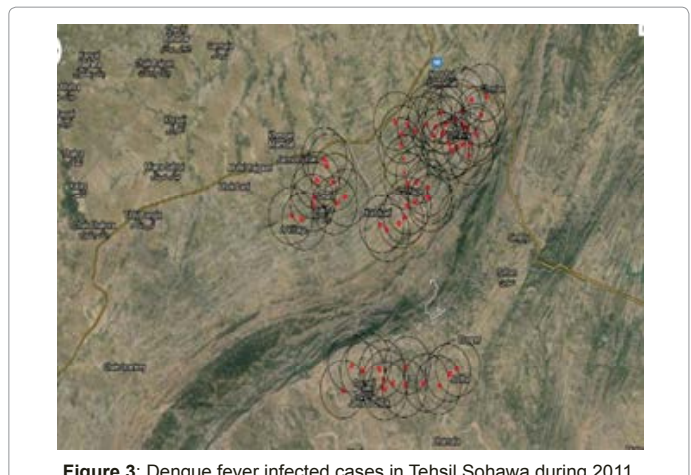

**Figure 3**: Dengue fever infected cases in Tehsil Sohawa during 2011.

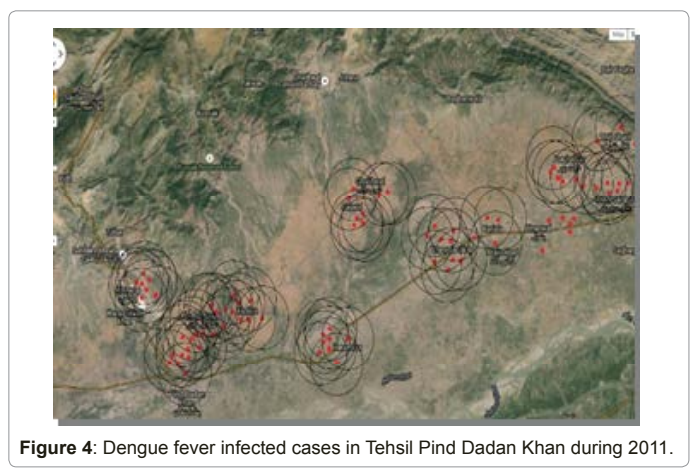

**Figure 4**: Dengue fever infected cases in Tehsil Pind Dadan Khan during 2011.

## K-medoids

Since k-means is sensitive to outliers and noisy data, so to minimize the sensitivity of k-means to outliers, k-mediod algorithm is proposed. To represent the clusters, k-mediod picks the actual objects instead of mean values' mediod is more resistant to outliers and noise. Since noise and outliers has less effect on mediods as compared to means that is why k-mediods works faster than k-means in the presents of outliers [17]. But k-mediods also has some drawbacks. It is more costly than k-means because it takes too much time to execute. It has to compute all pairwise distances. It does not work well with large set of data. Like k-means, user has to predict k-value in k-mediods also. Selection of k-value is also an issue. Although it is resistant to outliers but some experiments have shown that k-mediods can also falsely identify the clusters due to the noise and outliers. Since each attribute has the same significance so we never know which attribute has participated more than others shown in (Figure 7) [18].

## DBSCAN

To remove the drawbacks of k-mediods, dbscan algorithm was proposed. It was the first density-based algorithm. It resolves many drawbacks of k-means and k-mediods. Dbscan algorithm can handle clusters of different shapes, size and density. It is also resistant to noise. By using Dbscan one can differentiate between a low density cluster and a denser cluster. Dbscan provide more comprehensive view of data. Dbscan algorithm just need two parameters for execution i.e. ε and MinPts. Dbscan is not sensitive to the ordering of points. But it cannot handle changing densities. It shows disturbed results when values of parameters are changed. Dbscan is sensitive to parameter values. Dbscan has a high run time complexity shown in (Figure 8) [19]. Dbscan showed good clusters because it is showing separate and distinct clusters for each tehsil. So a better visualization of our dataset is achieved.

## Optics

Optics is an evolution of dbscan algorithm. Optics is the most sophisticated approach in the clustering techniques. Unlike dbscan, it shows comparatively large resistance to the change in parameter values. Optics is not sensitive to the distance and parameters. Optics can identify that data having high density which is located in low density groups. Clustering order do not matter in clustering results when using Optics [20]. Final clusters are resistant to parameters shown in (Figure 9).

## Comparison

From the (Figure 6), we can observe that k-means is not showing correct results because according to our dataset it should differentiate between different areas. But in this figure, k-means is overlapping tehsil Dina and Sohawa. When a viewer see the (Figure 6) he can only see three clusters. He cannot efficiently describe the distinction of dengue attack in different areas. There should be some distinction for these both tehsils. So this is not a good visualization. In the (Figure 7) although k-mediods resolved the issue raised in k-means to some extent. From the (Figure 7) we can analyze distinct clusters for each tehsil but still these clusters are not satisfactory. From (Figure 8) we can observe that dbscan has clearly discriminate between different tehsils. It is showing a satisfactory visualization of dengue attack. But still it is not providing detailed overview of our dataset i.e. there are many areas

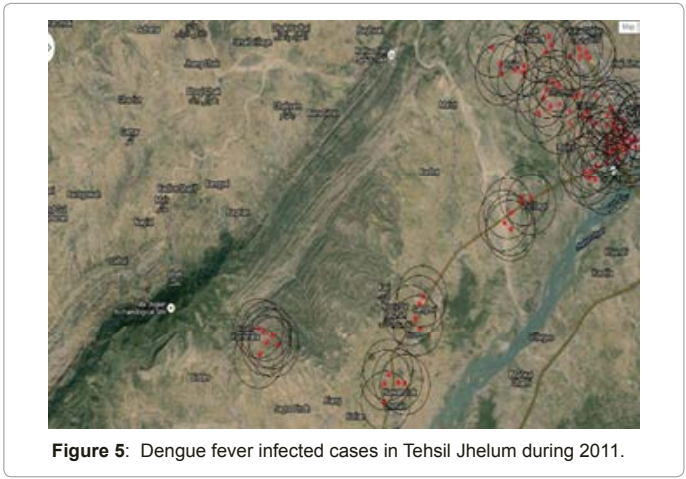
**Figure 5:** Dengue fever infected cases in Tehsil Jhelum during 2011.

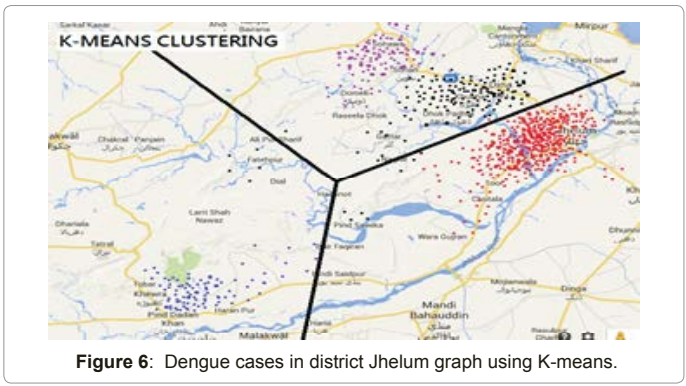
**Figure 6:** Dengue cases in district Jhelum graph using K-means.

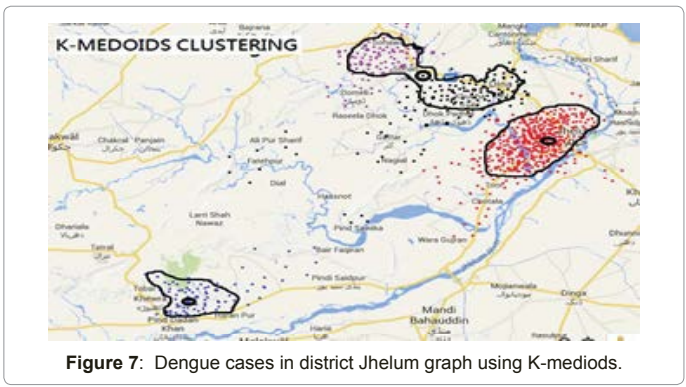
**Figure 7:** Dengue cases in district Jhelum graph using K-mediods.

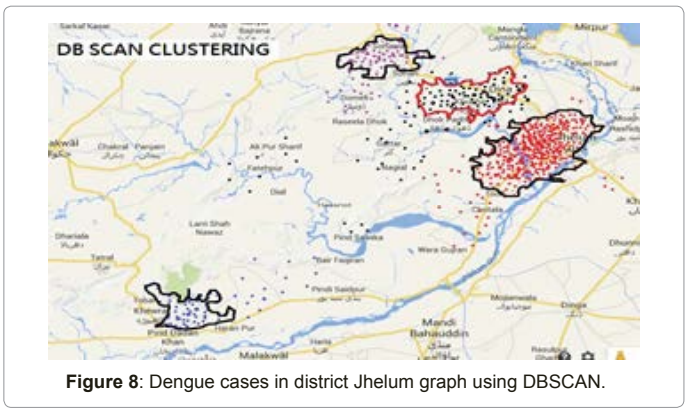
**Figure 8:** Dengue cases in district Jhelum graph using DBSCAN.

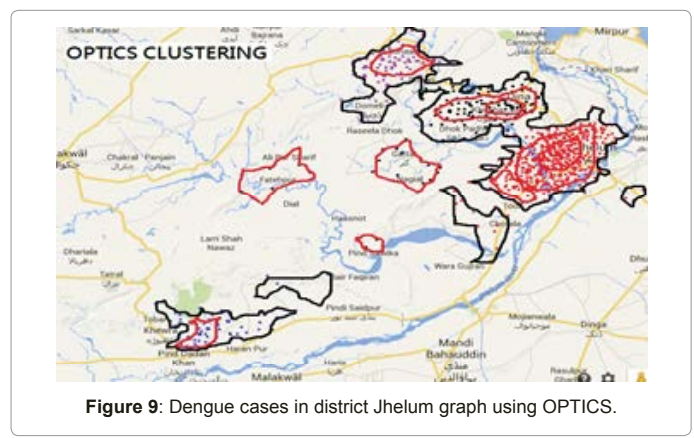

**Figure 9**: Dengue cases in district Jhelum graph using OPTICS.

at sub-tehsil level which should be declared as separate clusters because they had severe attack of dengue but dbscan didn't showed them as separate clusters. So this is a drawback of dbscan. So to achieve a detailed overview we'll apply Optics. And we can observe from (Figure 9) that optics is providing a really good and detailed visualization of our dataset. It has showed each area having dengue attack as separate cluster.

## Conclusion

Clustering techniques are very good tools for the visualization of diseases. And we have used density-based clustering methods for the application of chunk of data. The reason behind using density-based clustering is that density-based clustering separates regions having high density from the regions who have low density. So one can clearly understand the detailed chunk of data. When we get our dataset, by viewing it we predicted that dengue has attacked mostly in Jhelum city and Tehsil Jhelum. But when we applied the chunk of dataset we came to know that dengue has attacked the whole district very badly and in areas far from Jhelum too. Results by applying clustering techniques were quit opposite as we have expected. Since we have applied density-based clustering so we can say that by using density-based clustering techniques for the visualization purpose of any disease or disaster attack, preventions and precautions can be made in future and many lives can be saved. In this research we have used clustering techniques for the clustering of dengue fever in Jhelum. These techniques can be used in other areas so that prevention from any disaster can be planed and then we have graph compared different clustering algorithms. As we have mentioned in the paper that the dbscan algorithm also has some drawbacks so visualization using dbscan might not have a good impact so one can use optics algorithm for the clustering purpose.

## Acknowledgements

Authors of this research paper are thankful from core of their heart and express a heartfelt gratitude to the Executive District Officer (health) District Jhelum for his kind contribution by providing useful data